\documentclass[conference]{IEEEtran}
\IEEEoverridecommandlockouts
\usepackage{cite}
\usepackage{amsmath,amssymb,amsfonts}
\usepackage{algorithmic}
\usepackage{graphicx}
\usepackage{textcomp}
\usepackage{xcolor}
\usepackage{siunitx}
\usepackage[hidelinks]{hyperref}
\usepackage{url}
\usepackage{orcidlink}
\def\BibTeX{{\rm B\kern-.05em{\sc i\kern-.025em b}\kern-.08em
    T\kern-.1667em\lower.7ex\hbox{E}\kern-.125emX}}
\begin{document}

\title{Simulating Gadolinium-Induced Magnetic Field Variations for Temperature Sensing with Magneto-Mechanical Resonators}

\newcommand\blfootnote[1]{%
  \begingroup
  \renewcommand\thefootnote{}\footnote{#1}%
  \addtocounter{footnote}{-1}%
  \endgroup
}

\author{\IEEEauthorblockN{Jonas Faltinath\IEEEauthorrefmark{1}\IEEEauthorrefmark{2}\,\orcidlink{0009-0003-4128-2948},
Miriam Schmitz\IEEEauthorrefmark{1}\IEEEauthorrefmark{2}, Fynn Foerger\IEEEauthorrefmark{1}\IEEEauthorrefmark{2}\IEEEauthorrefmark{3}\,\orcidlink{0000-0002-3865-4603}, Martin Möddel\IEEEauthorrefmark{1}\IEEEauthorrefmark{2}\,\orcidlink{0000-0002-4737-7863} and
Tobias Knopp\IEEEauthorrefmark{1}\IEEEauthorrefmark{2}\IEEEauthorrefmark{3}\,\orcidlink{0000-0002-1589-8517}}
\IEEEauthorblockA{
Email: jonas.faltinath@tuhh.de \\
\IEEEauthorrefmark{1}Section for Biomedical Imaging, University Medical Center Hamburg-Eppendorf, Hamburg, Germany\\
\IEEEauthorrefmark{2}Institute for Biomedical Imaging, Hamburg University of Technology, Hamburg, Germany\\
\IEEEauthorrefmark{3}Fraunhofer Research Institution for Individualized and Cell-based Medical Engineering IMTE, Lübeck, Germany
}}

\maketitle

\begin{abstract}
Small-size magneto-mechanical resonators (MMR) represent an emerging class of passive, wireless sensors that combine a sensing functionality with a tracking option. The operation principle is based on a resonating rotor oscillation whose frequency is defined by the magnetic flux density of a stator magnet.
One general sensing mechanism is the coupling of an external parameter to this resonator frequency.
In this study, we investigate an approach for encoding a temperature information as a shift in the natural oscillation frequency utilizing the temperature-dependent magnetic properties of gadolinium (Gd).
We perform an isolated simulation study 
on the temperature scaling of the magnetic field generation for stators coated with Gd of varying thickness.
Our results show that the magnetic phase transition of Gd at its Curie temperature leads to a pronounced change in the magnetic permeability enabling a significant magnetic shielding behavior only for lower temperatures. In the transition regime, we find a peak sensitivity reaching $\SI{45.8}{\Hz / \kelvin}$ exceeding existing values from the literature by up to a factor of $\sim 20$.
The findings of this work are an important step toward quantitative high-sensitivity temperature extraction with MMRs.

\end{abstract}


\section{Introduction}
\begin{figure}[t]
\centerline{\includegraphics[width=0.65\columnwidth]{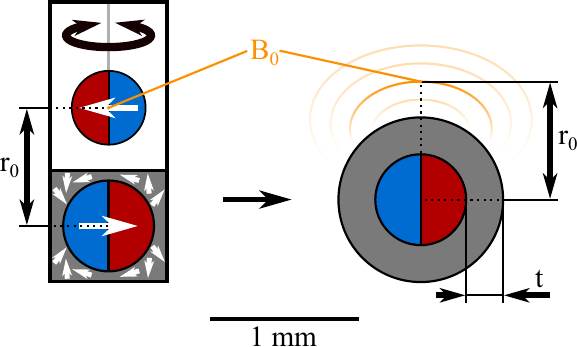}}
\caption{Left: Schematics of a temperature-responsive MMR. The stator magnet is attached to the sensor housing, while the rotor magnet is suspended by a filament (white arrows mark the magnetic moment direction). After excitation, the rotor performs a damped torsional oscillation which is driven by the magnetic interaction of both magnets. Assuming a magnetic dipole interaction, the stator magnetic flux density $B_0 = B(r_0)$ at the rotor center position $r_0$ is the primary contributor to the oscillation frequency. The incorporation of a material with temperature-dependent magnetic behavior (gray) like gadolinium (Gd) leads to a variation of $B_0$ and, correspondingly, to a temperature-sensitive frequency response. Right: Model system of this study. We assume a uniform Gd coating with thickness $t$ to simulate the temperature dependence in the magnetic flux density generation (orange).}
\label{fig:components}
\end{figure}

\blfootnote{\noindent This work was supported by the Deutsche Forschungsgemeinschaft (DFG, German Research Foundation), SFB 1615, Project number 503850735 and 555456057.} 
The precise knowledge of a system's temperature is crucial in a large number of contexts as it directly impacts e.g. chemical reaction yields and kinetics~\cite{smith2008reaction,blandamer1982dependence}, cell growth~\cite{knapp2022effects},
or the mechanism and success of thermal medical treatments~\cite{rossmann_review_2014,chu_thermal_2014,bianchi_thermophysical_2022,geoghegan2022monitor}.
This highlights the need for fast, sensitive, and accurate temperature sensors across various scientific and practical fields.
In this work, we investigate a novel temperature sensing approach based on miniature magneto-mechanical resonators (MMR). The main components of the MMR sensor are illustrated in Fig.~\ref{fig:components}. In contrast to state-of-the-art measurement devices like resistive sensors, thermocouples, acoustic waveguides, or fiber optic thermometers~\cite{Chen23}, the MMR sensors benefit from being entirely wireless and passive. The MMR platform also allows for a mapping of temperature values to spatial positions by an inherent localization capability. This provides an advantage compared to remote thermal imaging or laser temperature sensors which provide only a surface information~\cite{Chen23}. The mapping is particularly useful for integrated sensors in moving systems (e.g. thermal ablation needles) or for temperature profiling in large volumes (e.g. in reactors).

The initial MMR publication of Gleich et al.~\cite{gleich2023miniature} described the general operation principle of the platform.
In short, an external magnetic field induces a deflection of the rotor from the antiparallel magnetic moments equilibrium alignment. Because of the magnetic restoring torque resulting from the stator magnetic flux density $B$ and the low damping constant, the rotor will oscillate around the filament with a characteristic angular frequency. 
The corresponding signals caused by the rotating magnetic moment can be inductively captured by a coil array and be used for tracking~\cite{gleich2023miniature}.
The approach to make the sensor sensitive to the environment temperature $T$ is to make $B$, and correspondingly the frequency, a function of the temperature. One realization uses the thermal expansion of the sensor housing and filament~\cite{gleich2023miniature} leading to an altered magnet distance that, in turn, results in a corresponding change of $B$~\cite{faltinath2025natural}. In principle, this approach is inherently available, but it comes with the drawback of rather low sensitivities. 

In this study, we address this limitation by using a new sensor concept that is based on the integration of an additional magnetic component in the sensor housing that contributes to $B$. This element is made of a material that features strongly temperature-dependent magnetic properties as they are typically found around the Curie point $T_C$ of a ferromagnetic-to-paramagnetic phase transition. 
To investigate the expected thermal response, we perform simulations for the magnetic field variation at different temperatures of stator magnets coated with the low-$T_C$ material gadolinium (Gd). We elaborate on the underlying working principle and evaluate the sensitivity of different design settings.

\section{Materials and Methods}
\subsection{Model System and Data Basis}
For a quantitative comparison of our approach with a thermal expansion, we design our model system to match the experimental conditions of \cite{gleich2023miniature}. The MMRs in that study consist of spherical magnets, a stator of diameter $d_\text{stat}=\SI{0.63}{\mm}$ and a rotor with $d_\text{rot}=\SI{0.5}{\mm}$, made of neodymium-iron-boron (NdFeB) and resonate at a frequency of $2\pi\times\SI{2}{\kilo \hertz}$. 
Given these specifications, we introduce several further assumptions.
We consider equality of the specified resonance frequency with the natural frequency of the oscillator which is a valid approximation in the case of small deflection angles and low damping~\cite{faltinath2025natural}. Furthermore, we assume a mass density of the magnets of $\rho = \SI{7600}{\kilogram/\meter^{3}}$ and a degree of magnetization of N52. The latter is considered to provide a homogeneous remanence magnetic flux density of $B_r=\SI{1.44}{\tesla}$. Under these conditions, the proposed dipole model from \cite{gleich2023miniature, KnoppMohnFoergerThiebenHackelbergFaltinathTsandaBobergMöddel} allows us to calculate the expected magnet center-to-center distance $r_0 = \SI{0.83}{\mm}$. In our study, we define this distance above the stator center as the reference position for a point-like rotor obeying the magnetic moment $m = \SI{7.5e-5}{\ampere \,\meter^2}$ and the moment of inertia $I = \SI{1.2e-14}{\kilogram\,\meter^2}$ of the extended magnet. Assuming these rotor properties as fixed, the dipole model gives a direct relation between the generated stator magnetic flux density $B_0 = B(r_0)$ and the natural angular frequency of the rotor by $\omega_\text{nat} = \sqrt{m\,B_0/I}$~\cite{gleich2023miniature}.

Consequently, we can reduce the complexity of an MMR by only simulating the subsystem responsible for the generation of the $B_0$-field at different temperatures but still evaluate the expected frequency response of a full sensor. In this work, our model system consists only of the stator magnet with the above mentioned geometry and material properties and its temperature-sensitive coating embedded in an air volume (relative permeability $\mu_r = 1$) with a diameter of $\SI{7.3}{\mm}$.
We assume the stator magnet to be fully encapsulated by a Gd coating of uniform but variable thickness $t = \SIrange{50}{250}{\micro \meter}$ (see Fig.~\ref{fig:components}). The utilized temperature-dependent magnetization characteristics of Gd are based on a linear interpolation of the dataset from Liu et al.~\cite{liu_high-performance_2023}. 
Accordingly, we obtain $B=B(H,T)$ curves (with the magnetic field $H$) in the range of $T = \SIrange{276}{369}{\kelvin}\,(\sim \SIrange{3}{96}{\celsius})$ around the Curie point of $T_C = \SI{292}{\kelvin}\,(\sim\SI{19}{\celsius})$.

To allow a quantitative performance evaluation, we derive the sensitivity of each setup as the change in the natural frequency per unit temperature $\partial \omega_\text{nat}/\partial T$ via an averaged two-sided numerical differentiation. We extract the temperature range in which our obtained sensitivities dominate compared to the magnitude of the thermal expansion sensitivity from \cite{gleich2023miniature} where $\partial \omega_\text{nat} / \partial T = -2\pi\times\SI{2.3}{\Hz/\kelvin}$ is reported. For a proper comparison, we calculate the mean value inside this high-sensitivity range by a numerical integration.

\subsection{Physical Background and Simulation Environment}
The objective of this study is to simulate the magnetic flux density profile of our model system with a particular focus on the field $B_0$ at the reference position $r_0$ in different thermal conditions.
We consider our setup to be free of currents and to fulfill the magnetostatic approximation. In this case, we can derive the magnetic field $\mathbf{H}$ from a magnetic scalar potential $V_m$ with $\mathbf{H} = -\nabla V_m$. We use the finite element method (FEM) to determine the spatial distribution of the magnetic flux density $\mathbf{B}$ using Gauss' law for magnetism $\nabla \cdot\mathbf{B} = 0$ by appropriate adjustment of $V_m$~\cite{comsol_acdc_2024}.
This is facilitated by the \textit{Magnetic Fields, No Currents} interface of the software COMSOL Multiphysics\textsuperscript{\textregistered}\,6.2 (COMSOL AB, Sweden).

Each of the three considered material domains (NdFeB, Gd and air) is modeled with a distinct magnetization model to describe the relationship between $\mathbf{H}$ and $\mathbf{B}$~\cite{comsol_acdc_2024}. Inside the stator magnet, we assume the fixed remanent flux density $B_r$, the Gd domain is described by the custom temperature-dependent $B(H,T)$-curve and inside the air domain the relationship is modeled via the relative permeability $\mu_r$. 
In order to save computational time, we exploit the symmetry of the setup and numerically simulate only one eighth of the system. Furthermore to suppress edge effects, we define the outermost part of the air as an infinite element domain. We then obtain the flux density values in the full volume by performing an antisymmetric mirroring across the normal plane of the stator magnetic moment direction and symmetric mirroring across the two orthogonal planes perpendicular to the normal plane.

\section{Results}
\begin{figure}[t]
\centerline{\includegraphics[width=\columnwidth]{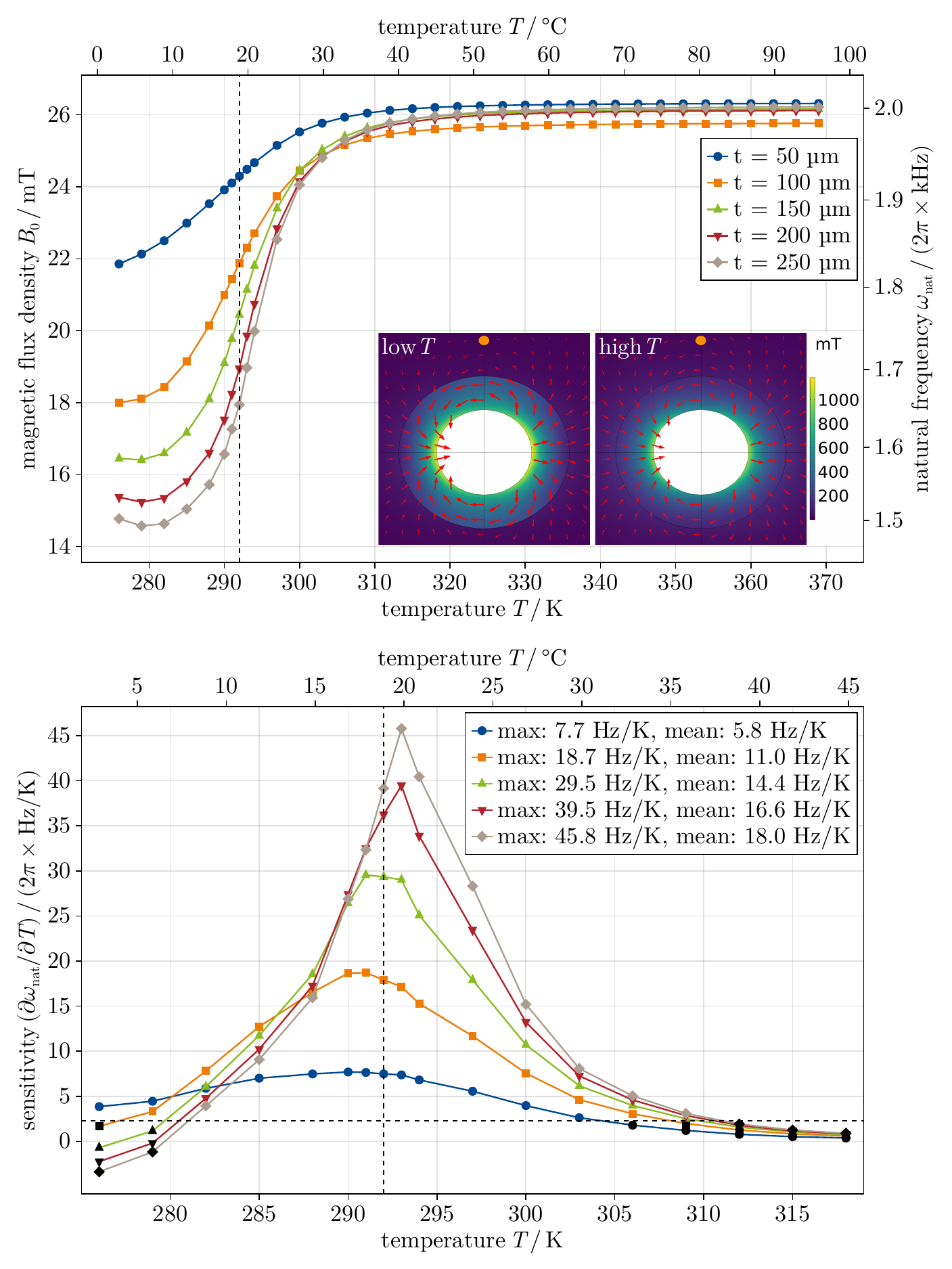}}
\caption{Magnetic flux density, natural frequency (top) and sensitivity (bottom) at the reference point $r_0$ for different temperatures and coating thicknesses. The peak sensitivities (max) are found close to the Curie temperature (vertical dashed line). For the mean sensitivities, the markers (black) below the thermal expansion sensitivity of Gleich et al.~\cite{gleich2023miniature} (horizontal dashed line) are neglected. The solid lines are a guide to the eye. Note the difference in the temperature axes between both plots.
The inset illustrates for $t = \SI{250}{\micro \meter}$ the pronounced magnetic shielding of $r_0$ (orange dot) due to the Gd coating at low $T$ ($\SI{282}{\kelvin}$). The shielding becomes weaker for high $T$ ($\SI{309}{\kelvin}$).}
\label{fig:results}
\end{figure}
We summarize and compare the simulation results in Fig.~\ref{fig:results}.
At low temperatures within the investigated range, the Gd coating leads to a decrease in the magnetic flux density to which the rotor would be exposed. Consequently, we also find a reduction of the natural frequency while for higher temperatures $\omega_\text{nat}$ converges toward the value expected for the bare stator of $2\pi\times\SI{2}{\kilo\hertz}$. Notably, the latter is independent of the coating thickness $t$. Due to the non-linear $B(T)$ profile, the sensitivities themselves are temperature-dependent. Hence, we find for each $t$ an individual and limited \textit{high-sensitivity \mbox{temperature} range}. 
Inside this interval, the Gd-induced sensitivity is larger than
the expected thermal expansion sensitivity~\cite{gleich2023miniature}. We highlight that all the corresponding ranges and sensitivity peaks are clustered around the Curie point. 

Our findings show a clear trend of a rising overall sensitivity with increasing coating thickness. As such, $t = \SI{250}{\micro \meter}$, which corresponds to a thickness-to-radius ratio of $\sim 0.8$, shows the best performance in our comparative study. We find a peak sensitivity of $\SI{45.8}{\Hz / \kelvin}$ at $T=\SI{293}{\kelvin}$ $(\sim\SI{20}{\celsius})$ and a mean sensitivity of $\SI{18.0}{\Hz / \kelvin}$ in a high-sensitivity range of $\SI{27}{\kelvin}$ located between $\SIrange{282}{309}{\kelvin}$ \mbox{$(\sim \SIrange{9}{36}{\celsius})$}. For this configuration, the inset of Fig.~\ref{fig:results} displays the 
local increase of the magnetic flux density and significant field line return inside the Gd coating only for low $T$. This leads to a peak-to-peak variation in $B_0$ by more than $\SI{11.6}{\milli \tesla}$ which corresponds to a shift larger than $\SI{500}{\Hz}$ or $\SI{25}{\percent}$ in the natural frequency. 

\section{Discussion}
Our results show that the incorporation of Gd as a passive sensing element into an MMR can significantly enhance its thermal sensitivity compared to a pure temperature-induced expansion. The obtained mean sensitivity of our best-performing setup with $t=\SI{250}{\micro \meter}$ exceeds reported values of other studies~\cite{gleich2023miniature,merbach2025wirelesspassivepressuredetection} by at least a factor of $\sim 8$, the peak value even by a factor of $\sim 20$. We note that a further increase in the coating thickness is impractical as it would result in a physical contact with the assumed rotor domain of our study. Importantly, the sensitivity sign within our high-sensitivity range is inverted compared to typical values from thermal expansions~\cite{gleich2023miniature,merbach2025wirelesspassivepressuredetection} leading in our case to larger natural frequencies if the temperature is increased.

This can be attributed to the fact that the Gd coating acts in our geometry as a temperature-dependent magnetic shield. At low $T$, it occurs in the ferromagnetic phase which exhibits a rather large relative magnetic permeability. Consequently, the material shunts the magnetic field lines from the stator~\cite{xu_magnetic_2023}, effectively reducing the flux density at the rotor location. This effect is even more pronounced for thicker shields because of the reduced magnetic reluctance or in multilayer setups because of parallel shunting~\cite{xu_magnetic_2023}. At higher temperatures, Gd transitions to a paramagnetic phase. Since its permeability then gradually approaches unity, the contribution to $B_0$ vanishes leaving only the undisturbed stator magnet field. Furthermore, the results of Fig.~\ref{fig:results} can be explained given that Gd has the steepest slope of its magnetic susceptibility in proximity to the Curie temperature. As the temperature response of our sensors scales with the variation, we also find the range of the highest sensitivity located around $T_C$. We note that, as the MMR working principle requires $B_0 \neq 0$, a full substitution of the stator magnet with a Gd element is not feasible. 

The magnetic shunting result depends on the precise coating arrangement. In a suitable design, it can even be used to guide the field lines for an intentional local enhancement in $B_0$ at the rotor position leading to a negative sensitivity. This flexibility opens the way toward tailored sensors, optimizing sensitivities or, from another perspective, even compensating for temperature drifts. The latter could be enabled by choosing the inverted scaling sign for the Gd-induced sensitivity compared to the inherent thermal expansion. This is particularly relevant to avoid cross-sensitivities in frequency-modifying sensing approaches of other quantities like pressure~\cite{merbach2025wirelesspassivepressuredetection}.

The presented sensor concept is not designed for universal temperature sensing. It exhibits a non-linear response function and the dynamic range is limited to the high-sensitivity window around the Curie point. Consequently, the peak performance for an implementation with Gd is found around room temperature. However, specific applications might require a distinct central temperature or set constraints on the width of the temperature range. The miniature MMR technology was initially developed for potential medical applications~\cite{gleich2023miniature} which e.g. benefit from an operation near body temperature. Future studies that focus on corresponding shifts in the operating point might involve the detailed analysis of different alloys of magnetic materials to provide an adequate $T_C$ tailoring~\cite{law_tunable_2010}.
In conclusion, our study demonstrates that leveraging the magnetic properties of Gd or similar materials presents a promising approach for designing  passive, wireless magneto-mechanical resonators with a strong thermal frequency response.

\bibliographystyle{IEEEtran}
\bibliography{ref}

\end{document}